\documentclass[conference,10pt]{IEEEtran}
\IEEEoverridecommandlockouts
\def\BibTeX{{\rm B\kern-.05em{\sc i\kern-.025em b}\kern-.08em
    T\kern-.1667em\lower.7ex\hbox{E}\kern-.125emX}}

\usepackage[utf8]{inputenc}
\usepackage{cite}
\usepackage{amsmath,amssymb,amsfonts}
\usepackage{graphicx}
\usepackage{textcomp}
\usepackage[caption=false]{subfig}
\usepackage{epsfig}
\usepackage{float}
\usepackage{color}
\usepackage{balance}
\usepackage{bbm}
\usepackage{textcomp}
\usepackage{enumitem}
\usepackage{comment}

\usepackage{mathtools,nccmath,amsthm}
\usepackage{multicol,tikz,array,booktabs,url}
\usepackage{xcolor}
\usepackage{algorithm}
\usepackage{algpseudocode,algorithmicx}
\usepackage{acronym}
\usepackage{xfrac}

\graphicspath{{figuresSPAWC}{./}}
\DeclareGraphicsExtensions{.pdf}

\newcommand{\vect}[1]{\boldsymbol{\mathbf{#1}}}

\DeclareMathAlphabet{\mathpzc}{OT1}{pzc}{m}{it}

\algnewcommand{\LineComment}[1]{\State \(\#\) #1}
\algnewcommand\algorithmicinput{\textbf{Set}}
\algnewcommand\Set{\item[\algorithmicinput]}
\algnewcommand\algorithmicinitial{\textbf{Initialize}}
\algnewcommand\Initialize{\item[\algorithmicinitial]}

\let\oldReturn\Return
\renewcommand{\Return}{\State\oldReturn}

\begin{document}
%\title{Robust Beamforming for Cell-Free OFDM Systems with Waveguide-Fed Dynamic Metasurface Arrays}
\title{Robust Beamforming for Cell-Free Systems with Parallel-Plate-Waveguided Dynamic Metasurfaces}

\author{Konstantinos D. Katsanos, Panagiotis Gavriilidis, and George C. Alexandropoulos\\
Department of Informatics and Telecommunications, National and Kapodistrian University of Athens, Greece
\\e-mails: \{kkatsan, pangavr, alexandg\}@di.uoa.gr
}
%\author{Konstantinos D. Katsanos$^1$ and George C. Alexandropoulos$^{1,2}$\\
%$^1$Department of Informatics and Telecommunications, National and Kapodistrian University of Athens, Greece
%\\$^2$Department of Electrical and Computer Engineering, University of Illinois Chicago, IL, USA
%\\e-mails: \{kkatsan, alexandg\}@di.uoa.gr
%}

\maketitle
\thispagestyle{empty}

\begin{abstract}
Dynamic Metasurface Antennas (DMAs) constitute a promising solution for extremely large antenna arrays, requiring lower power consumption and reduced hardware cost as compared to conventional phased arrays. In this paper, we consider a cell-free Orthogonal Frequency Division Multiplexing (OFDM) system comprising multiple Base Stations (BSs) equipped with parallel-plate-waveguided DMAs, which aims to serve multiple users in the downlink direction. Focusing on a realistic frequency-selective model for the response-tunable elements of each DMA panel, and targeting to surpass the necessity of centralized designs that rely on a central processing unit with high computational power, we present a distributed optimization framework with minimal control information exchange for the frequency-selective analog and digital beamforming matrices of the multiple BSs, having the system spectral efficiency maximization as the design objective. Considering imperfect Channel State Information (CSI) availability at each BS, we devise a parallel decomposition framework for the configuration of the tunable parameters of each DMA-based BS. Our numerical results showcase the robustness of the proposed distributed beamforming design over different CSI conditions, and quantify the critical role of taking into account mutual coupling during the DMA design process.
\end{abstract}


\section{Introduction} 
The upcoming sixth Generation (6G) of wireless networks is expected to leverage metasurface-based antenna technology to realize eXtremely Large (XL) arrays \cite{DMA_Nir2021}. These architectures offer a scalable solution, where increasing the number of metamaterial elements enhances directivity and spatial degrees of freedom without a proportional increase in power consumption. This is due to the fact that signal manipulation is primarily performed in the analog domain, reducing the need for large chunks of Radio-Frequency (RF) chains.

A prominent implementation of such architectures is the Dynamic Metasurface Antenna (DMA), typically realized using planar structures formed by stacking microstrip lines~\cite{DMA_Nir2021,williams2023EM_DMA}. Each line is individually fed and loaded with reconfigurable metamaterial elements. In this paper, we instead consider a DMA architecture based on a 2D waveguide, namely the Parallel Plate Waveguide (PPW) \cite{pulidomancera2018}. Unlike microstrip-based designs, the PPW constitutes a truly planar structure that does not rely on stacking linear arrays, and can naturally scale to XL apertures by increasing the plate dimensions. Importantly, this does not require per-line feeding and can, in principle, operate with a single RF chain. Additionally, the PPW supports the fundamental TEM mode, which has no lower cutoff frequency, leading to more favorable broadband characteristics~\cite{davidsmith2017}.

Concurrently, the cell-free network paradigm has gained substantial interest in recent years, mainly due to its potential to mitigate the inherent limitations of conventional cellular architectures~\cite{he2021cell}. Nonetheless, typical cell-free deployments rely on Central Processing Units (CPUs) with high computational capabilities, which coordinate transmissions over dedicated backhaul links to the distributed Base Stations (BSs)~\cite{ngo2017cell}. Very recently, in~\cite{du2025joint}, the use of DMAs within cell-free systems was introduced, demonstrating their advantages for joint user association and efficient transmit beamforming, when the design objective is the weighted sum-rate maximization. In addition, the importance of partially decentralized strategies to alleviate the computational burden at the CPU was highlighted.

This paper investigates a cell-free Orthogonal Frequency Division Multiplexing (OFDM) system with multiple (BSs), each equipped with a PPW-based DMA, that jointly serves several single-antenna User Equipments (UEs) in the downlink. Explicitly accounting for the frequency-selective DMA behavior, we develop a distributed design of the analog and digital beamforming matrices under imperfect Channel State Information (CSI) conditions. Simulation results demonstrate the robustness of the proposed design, and highlight the critical role of incorporating mutual coupling in the DMA design. 

%----------------------------------------------------------%
%                      Section Change                      %
%----------------------------------------------------------%
\section{PPW-Based DMA and Channel Modeling}\label{sec:TX_modeling}
Each multi-antenna BS in the considered cell-system is equipped with a DMA realized through a PPW of height $h$, filled with air \cite{pulidomancera2018}. The metasurface is particularly excited by multiple feeds placed inside the waveguide, and radiates through subwavelength metamaterial elements etched on the upper plate. Due to their electrically small dimensions, these elements can be accurately described using the discrete dipole approximation~\cite{williams2023EM_DMA,pulidomancera2018,davidsmith2017}. To this end, the response of each metamaterial element is described by a polarizability tensor that relates the local magnetic field to the induced magnetic dipole moment. However, for the class of slot-based elements considered in this paper, the dominant response occurs along a single direction. Consequently, each element is characterized by a scalar polarizability coefficient, and we assume the dominant axis of the metamaterial to be along $x$, while the metasurface is assumed to lie on the $x$-$y$ plane at $z=0$. 

Let $m_n\in\mathbb{C}$ denote the magnetic dipole moment of each $n$-th element $(n=1,2,\ldots,N$, where $N$ is the total number of elements) and $H_{\text{loc},n}\in\mathbb{C}$ the local $x$-directed magnetic field at its location. Their relation is expressed as $m_n = \alpha_n H_{\text{loc},n}$, 
%\begin{equation}\label{eq: magnetic moment from local field_scalar}
%    m_n = \alpha_n H_{\text{loc},n},
%\end{equation}
where $\alpha_n$ denotes the scalar magnetic polarizability of the $n$-th element. Although the polarizability is generally frequency dependent, i.e., $\alpha_n=\alpha_n(f)$, the explicit dependence on $f$ is omitted for notational simplicity. The local magnetic field at each element consists of two contributions: the guided excitation generated by the feeds and the scattered fields produced by the remaining dipoles in the array, hence:
\begin{equation} \label{eq:Hloc_scalar}
    H_{\text{loc},n} = H_{0,n} + \sum_{j=1,\,j\neq n}^{N} G_{n,j}\, m_j,
\end{equation}
where $H_{0,n}$ represents the magnetic field induced by the feeds at the $n$-th element location, and $G_{n,j}$ models the electromagnetic interaction between elements $j$ and $n$, excluding the self-interaction term. Let $\mathbf{r}_n \triangleq [r_{x_n},r_{y_n},0]^{\rm T}$ denote the position of the $n$-th element. The interaction coefficients are given for $n\neq j$ as: $G_{n,j}\triangleq G_{\text{WG}}(\mathbf{r}_n-\mathbf{r}_j)+G_{\text{FS}}(\mathbf{r}_n-\mathbf{r}_j)$, and zero for $n=j$. In particular, $G_{\text{WG}}$ captures the coupling through the waveguide and $G_{\text{FS}}$ accounts for the coupling through free-space radiation, and they are given in \cite[eqs.~(6) and (7)]{Gavriilidis2026WaveguideMetasurface}. 

\begin{comment}
For a PPW of height \(h\), the waveguide interaction between two \(x\)-directed dipoles is given by \cite{pulidomancera2018}:
\begin{equation}
\begin{split}
G_{\text{WG}}(\mathbf{r}_n-\mathbf{r}_j)
    = -\frac{\jmath k^2}{8h}\Big[
        H^{(2)}_{0}\!\left(k\rho_{n,j}\right)
        - \cos(2\psi_{n,j})H^{(2)}_{2}\!\left(k\rho_{n,j}\right)
    \Big],
\end{split}
\label{eq:WaveguideGreenScalar}
\end{equation}
where \(H^{(2)}_{\nu}(\cdot)\) denotes the Hankel function of the second kind and order \(\nu\) \cite{balanis2016antenna}, \(k\) is the propagation constant of the guided mode, \(\rho_{n,j}\triangleq\lvert\mathbf{r}_n-\mathbf{r}_j\rvert\), and \(\psi_{n,j}\triangleq\mathrm{atan}\left((r_{y_n}-r_{y_j})/(r_{x_n}-r_{x_j})\right)\).

The free-space contribution follows from the magnetic dipole dyadic Green's function \cite{Novotny_Hecht_2006}. Retaining only the \(x\)-to-\(x\) interaction component and accounting for the image dipole created by the metallic plate leads to \cite{gavriilidis2026nearfield}:
\begin{equation}
\begin{split}
G_{\text{FS}}(\mathbf{r}_n-\mathbf{r}_j)
    = \Bigg[
    \left(\frac{3}{k^2 \rho_{n,j}^2}+\frac{3\jmath}{k\rho_{n,j}}-1\right)\cos^2(\psi_{n,j})\\
    +\left(1-\frac{\jmath}{k\rho_{n,j}}-\frac{1}{k^2 \rho_{n,j}^2}\right)
    \Bigg]\frac{k^2 e^{-j k \rho_{n,j}}}{2\pi \rho_{n,j}} .
\end{split}
\label{eq:FreeSpaceGreenScalar}
\end{equation}
\end{comment}

The feeds inside the PPW of each DMA are modeled as independently driven thin-wire current sources placed at positions $\mathbf{p}_i\triangleq[p_{x_i},p_{y_i},0]^{\rm T}$ $\forall i=1,\ldots,N_f$. The magnetic field at the $n$-th element due to all feeds is obtained by superposing the contributions of each source \cite{Gavriilidis2026WaveguideMetasurface}, as follows:
\begin{equation} \label{eq:H0_scalar}
   \!\!\!\! H_{0,n}\! \triangleq\!
    \frac{\jmath \beta}{4}
    \sum_{i=1}^{N_f} I_i H^{(2)}_{1}\!\left(\beta|\mathbf{r}_n-\mathbf{p}_i|\right)
    \sin\!\!\left(\!\!
    \mathrm{atan}\left(\!\frac{r_{y_n}-p_{y_i}}{r_{x_n}-p_{x_i}}\!\right)
    \!\!\right)\!\!,\!\!\!
\end{equation}
where $H^{(2)}_{1}(\cdot)$ denotes the Hankel function of the second kind and order $1$ \cite{balanis2016antenna}, while $\beta$ is the propagation constant of the guided mode. For compactness, we define $\mathbf{m} \triangleq [m_1,\ldots,m_N] \in \mathbb{C}^{N\times1}$, $\mathbf{A}\triangleq\mathrm{diag}[\alpha_1,\ldots,\alpha_N]\in\mathbb{C}^{N\times N}$, and the interaction matrix $\mathbf{G}\in\mathbb{C}^{N\times N}$ with entries $[\mathbf{G}]_{n,j}=G_{n,j}$. Likewise, the excitation field vector is defined as $\mathbf{h}_0 \triangleq [H_{0,1},\ldots,H_{0,N}] \in \mathbb{C}^{N\times1}$. Furthermore, we define the excitation matrix $\mathbf{H}_f\in\mathbb{C}^{N\times N_f}$, whose entries are
$[\mathbf{H}_f]_{n,i} = \frac{\jmath\beta}{4} H^{(2)}_{1}\!\left(\beta|\mathbf{r}_n-\mathbf{p}_i|\right) \sin\!\left( \mathrm{atan}\left(\frac{r_{y_n}-p_{y_i}}{r_{x_n}-p_{x_i}}\right)\right)$. Letting $\mathbf{i}\triangleq[I_1,\ldots,I_{N_f}] \in \mathbb{C}^{N_f\times1}$ contain the currents at the feeds, the excitation field can be written as $\mathbf{h}_0=\mathbf{H}_f\mathbf{i}$. Combining all the above, yields the dipole moment solution \cite{Gavriilidis2026WaveguideMetasurface}: 
\begin{equation} \label{eq:dipole_current_relation_scalar}
    \mathbf{m} = \underbrace{\left(\mathbf{A}^{-1}-\mathbf{G}\right)^{-1}}_{\triangleq \mathbf{W}_{\rm RF}} \mathbf{H}_f \mathbf{i},
\end{equation}
where $\mathbf{W}_{\rm RF}\in\mathbb{C}^{N\times N}$ is essentially the analog beamformer, while $\mathbf{i}$ bears the digital precoding and information symbol; the latter can thus be written as $\mathbf{i}=\mathbf{Vs}$, whereas $\mathbf{V}$ denotes the precoder and $\mathbf{s}$ the information symbol. Assuming a lossless and passive metamaterial element, the polarizability of the $n$-th element can be modeled as follows \cite[eq.~(49)]{Smith2025PolarizabilityandImpedance}:
\begin{equation}\label{eq:Lorentzian}
    \alpha_{n}(f) = \frac{\alpha_{0,n} f_{0,n}^2}{f_{0,n}^2 - f^2 + \jmath \alpha_{0,n} f_{0,n}^2 C(f)},
\end{equation}
where $f_{0,n}\in\mathbb{R}$ denotes the resonance frequency, $\alpha_{0,n}\in\mathbb{R}$ controls the resonance strength of the $n$-th element, and \(C(f)\triangleq \frac{8\pi^2 f^3}{3 c^3} + \frac{\pi^2 f^2}{2 h c^2}\) models radiation damping in the PPW, and follows from power conservation \cite{Gavriilidis2026WaveguideMetasurface}, with \(c\) being the speed of light. It is also noted that all matrices involved in \eqref{eq:dipole_current_relation_scalar} are frequency dependent. 

\vspace{-0.3cm}
\subsection{Free-Space Channel Model} \label{sec:channel_model}
The electric field radiated in free space can be expressed as a function of the magnetic dipole moments, via: $\mathbf{e}_{{\rm sc},n}(\mathbf{r}) \triangleq \jmath\omega\mu_0
\big[\nabla\times \tilde{\mathbf{G}}_{\rm FS}(\mathbf{r}-\mathbf{r}_n)\big] \mathbf{m}_n$, where $\tilde{\mathbf{G}}_{\rm FS}\in\mathbb{C}^{3\times3}$ denotes the free-space dyadic Green's function \cite[eq.~(8.61)]{Novotny_Hecht_2006}. The expression is doubled to account for the image dipole created by the metallic plate of the waveguide. By adopting the far-field approximation, let $\mathbf{s}_u$ denote the observation point. Taking the origin at the center of the transmit aperture, $R_u \triangleq \|\mathbf{s}_u\|$ denotes the distance from the aperture center to the observation point, and $(\theta_u,\phi_u)$ represents the corresponding elevation and azimuth angles. Moreover, letting $\hat{\mathbf{s}}_u \triangleq [\sin\theta_u\cos\phi_u,\,
\sin\theta_u\sin\phi_u,\,\cos\theta_u]^{\rm T}$ be the unit vector pointing from the transmit DMA center toward $\mathbf{s}_u$, then, the electric field radiated by the $n$-th dipole can be expressed as: 
\begin{align} \label{eq: electric_field_due_to_nth_dipole_scalar}
e_{{\rm sc},n}^{\theta}(\mathbf{s}_{u}) &= \frac{\eta \beta^{2}}{2\pi R_u} e^{-\jmath \beta R_u} e^{\jmath \beta \hat{\mathbf{s}}_u^{\rm T}\mathbf{r}_n} \, m_n \sin(\phi_{u}), \\
e_{{\rm sc},n}^{\phi}(\mathbf{s}_{u}) &= \frac{\eta \beta^{2}}{2\pi R_u} e^{-\jmath\beta R_u} e^{j \beta \hat{\mathbf{s}}_u^{\rm{T}}\mathbf{r}_n} \, m_n \cos(\phi_{u})\cos(\theta_{u}), 
\end{align}
where $\eta\triangleq120\pi$ denotes the free-space impedance. The electric-field components radiated toward the observation point $\mathbf{s}_u$ can then be written with respect to the stacked dipole moment vector $\mathbf{m}\in\mathbb{C}^{N\times1}$ as: $e_{\rm sc}^{\theta}(\mathbf{s}_u)=\mathbf{h}_\theta^{\rm T}(\mathbf{s}_u)\mathbf{m}$ and $e_{\rm sc}^{\phi}(\mathbf{s}_u)=\mathbf{h}_\phi^{\rm T}(\mathbf{s}_u)\mathbf{m}$, where $\mathbf{h}_\theta(\mathbf{s}_u),\mathbf{h}_\phi(\mathbf{s}_u)\in\mathbb{C}^{N\times1}$ collect the contribution of each metasurface element toward the observation direction. Their $n$-th entries are given as follows:
\begin{align}
[\mathbf{h}_\theta(\mathbf{s}_u)]_n &= \frac{\eta \beta^{2}}{2\pi R_u} e^{-\jmath\beta R_u} e^{\jmath\beta \hat{\mathbf{s}}_u^{\rm T}\mathbf{r}_n} \sin(\phi_u),\\
[\mathbf{h}_\phi(\mathbf{s}_u)]_n &= \frac{\eta \beta^{2}}{2\pi R_u} e^{-\jmath\beta R_u} e^{\jmath\beta \hat{\mathbf{s}}_u^{\rm T}\mathbf{r}_n} \cos(\phi_u)\cos(\theta_u).
\end{align}

Assume that each BS serves $U$ UEs, where each $u$-th UE ($u=1,\ldots,U$) is modeled as an ideal thin-wire dipole of length $\ell_u$ with orientation unit vector $\hat{\mathbf{d}}_u\triangleq[\sin\vartheta_u\cos\varphi_u,\sin\vartheta_u\sin\varphi_u,\cos\vartheta_u]^{\rm T}$. Then, the received signal, which corresponds to the open-circuit voltage induced along the wire, is given by the line integral of the electric field along the dipole. By assuming that the electric field remains approximately constant over the antenna length, the induced voltage is approximated as $\ell_u\hat{\mathbf{d}}_u^{\rm T}\mathbf{e}$, where we use the spherical decomposition $\mathbf{e}\triangleq e^\theta\hat{\boldsymbol{\theta}}_u+e^\phi\hat{\boldsymbol{\phi}}_u$, with $\hat{\boldsymbol{\theta}}_u\in \mathbb{R}^{3 \times 1}$ and $\hat{\boldsymbol{\phi}}_u\in \mathbb{R}^{3 \times 1}$ being the spherical polarization unit vectors. This leads to the polarization projection coefficients $\gamma_{u,\theta}\triangleq\hat{\mathbf{d}}_u^{\rm T}\hat{\boldsymbol{\theta}}_u$ and $\gamma_{u,\phi}\triangleq\hat{\mathbf{d}}_u^{\rm T}\hat{\boldsymbol{\phi}}_u$. Consequently, the polarization-projected channel, that maps from metasurface dipole moments to voltage at each $u$-th UE, is expressed as follows:
\begin{equation}\label{eq:channel_matrix}
\mathbf{h}_u \triangleq \ell_u \left(\gamma_{u,\theta}\mathbf{h}_\theta(\mathbf{s}_{u}) + \gamma_{u,\phi}\mathbf{h}_\phi(\mathbf{s}_{u})\right)\in\mathbb{C}^{N\times1}.
\end{equation}
% Consequently, the received signal of user $u$ can then be written compactly as:
% \begin{equation}\label{eq: received signal}
% {y}_u = \mathbf{h}^{\rm T}_u \mathbf{W}_{\rm RF}\mathbf{H}_f
% \mathbf{i} + {n}_u,
% \end{equation}
% where \({n}_u\sim\mathcal{CN}({0},\sigma_u^2)\) models the AWGN, while \(\mathbf{h}_u\in\mathbb{C}^{N\times 1}\) is 
% It is noted that all matrices involved are frequency dependent. In the multicarrier formulation, they can therefore be indexed per subcarrier, e.g., \(\tilde{{y}}[t]\) denotes the received signal at the \(t\)-th subcarrier. Regarding the polarizability of each metasurface element, we adopt the Lorentzian model while accounting for the radiation damping specific to the PPW structure. For passive elements, the radiation damping term is given by \(C(f)=k^3/(3\pi)+k^2/(8h)\) \cite{gavriilidis2026nearfield}. 

%----------------------------------------------------------%
%                      Section Change                      %
%----------------------------------------------------------%
\section{System Model and Problem Formulation} 
In this section, we present the considered cell-free OFDM system comprising $B$ DMA-based BSs that concurrently serve $U$ single-antenna UEs in the downlink over a bandwidth ${\rm BW}$. We first present the received signal model and the adopted performance metric, and subsequently formulate the hybrid analog and digital beamforming design problem at the BSs. 

\vspace{- 0.3 cm}
\subsection{Received Signal Model} 
We assume that each $b$-th BS ($b=1,2,\dots,B$) is equipped with a PPW-based DMA that comprises \(N_f\) transmit RF chains and $N$ frequency-selective response-tunable metamaterial elements, whose configuration is capable of creating a resonant response. Each $b$-th BS communicates with each $u$-th UE over $K$ equally spaced SubCarriers (SCs) throughout the range ${\rm BW}$ around the carrier frequency $f_c$, precoding the transmit symbol at each $k$-th SC ($k=1,\ldots,K$), $\vect{s}[k]\triangleq[s_1[k],\ldots,s_U[k]]\in\mathbb{C}^{U\times 1}$ (with $\mathbb{E}[|s_u[k]|^2] = 1$ and $\mathbb{E}[s_u[k]s_q^*[k]] = 0$ $\forall q\neq u$), first, through the digital beamformer $\vect{V}_b[k]\in\mathbb{C}^{N_f\times U}$ and then via the analog beamformer $\vect{W}_{{\rm RF},b}[k]\in\mathbb{C}^{N\times N}$, as defined in \eqref{eq:dipole_current_relation_scalar}. In particular, the latter matrix is given for each $b$-th BS equipped a PPW-based DMA by the expression:
\begin{equation} \label{eqn:W_RF_definition}
\vect{W}_{{\rm RF},b}[k] \triangleq \left( \vect{A}_b^{-1}[k] - \vect{G}_b[k] \right)^{-1},
\end{equation}
where $\vect{A}_b[k] \triangleq \mathrm{diag}\left[\alpha_{b,1}(f_k),\alpha_{b,2}(f_k),\dots,\alpha_{b,N}(f_k)\right]$, with $\alpha_{b,n}(f_k)$ given by the Lorentzian expression in \eqref{eq:Lorentzian}, evaluated at the $k$-th SC's frequency. The transmit signal vector at each SC $k$ from each $b$-th BS can be mathematically expressed as:
\begin{equation} \label{eqn:transmit_signal}
    \vect{x}_b[k] \triangleq \vect{W}_{{\rm RF},b}[k] \vect{H}_{f,b}[k] \vect{V}_b[k]\vect{s}[k],
\end{equation}
where $\vect{H}_{f,b}[k]\in\mathbb{C}^{N\times N_f}$ represents the excitation matrix of the $b$-th DMA-based BS, as described in Section~\ref{sec:TX_modeling}. 
% , which is defined as follows:
% \begin{equation*}%\label{excitation_matrix}
% [\mathbf{H}_f]_{n,i}
% \triangleq
% \frac{\jmath k}{4}
% H^{(2)}_{1}\!\left(k|\mathbf{r}_n-\mathbf{p}_i|\right)
% \sin\!\left(
% \mathrm{atan}\left(\frac{p_{y_i}-r_{y_n}}{p_{x_i}-r_{x_n}}\right)
% \right).
% \end{equation*}

Let $\vect{h}_{b,u}[k]\in\mathbb{C}^{N\times 1}$ represent the channel gain vector for SC $k$ between each $b$-th DMA-based BS and each $u$-th UE. The baseband received signal at this SC at the $u$-th UE is ($n_u[k]\sim\mathcal{CN}(0,\sigma_{u,k}^2)$ is the additive white Gaussian noise):
\begin{equation} \label{eqn:received_signal_u_k}
\begin{aligned}
    &y_u[k] \triangleq \sum_{b=1}^B\vect{h}_{b,u}^{\rm H}[k]\vect{W}_{{\rm RF},b}[k]\vect{H}_{f,b}[k]\vect{v}_{b,u}[k]s_u[k] \\
    &+\sum\limits_{\substack{q=1,\\q\neq u}}^U\sum_{b=1}^B\vect{h}_{b,u}^{\rm H}[k]\vect{W}_{{\rm RF},b}[k]\vect{H}_{f,b}[k]\vect{v}_{b,q}[k]s_q[k] + n_u[k],
\end{aligned}
\end{equation}
where $\vect{v}_{b,u}[k]\in\mathbb{C}^{N_f\times 1}$ is the $u$-th column of $\vect{V}_b[k]$. 

\subsection{Design Problem Formulation} 
The achievable instantaneous sum-rate performance (measured in bits per second per Hz (bits/s/Hz)) for the considered multi-carrier cell-free system can be compactly expressed as a function of the $BK$ analog, $\widetilde{\vect{W}}_{\rm RF}\triangleq\{\vect{W}_{{\rm RF},1}[1],\ldots,\vect{W}_{{\rm RF},B}[K]\}$, and $BKU$ digital, $\tilde{\vect{v}}\triangleq\{\vect{v}_{1,1}[1],\ldots,\vect{v}_{B,U}[K]\}$, beamformers, as follows:
\begin{equation}
    \mathcal{R}\left(\widetilde{\vect{W}}_{\rm RF},\tilde{\vect{v}}\right)=\frac{1}{K}\sum_{u=1}^U\sum_{k=1}^K\log_2(1 + {\rm SINR}_{u,k}),
\end{equation}
with the Signal-to-Noise-plus-Interference Ratio (SINR) per $k$-th SC at each $u$-th UE included in this expression given by:
\begin{equation}
    {\rm SINR}_{u,k} = \frac{\left\lvert \sum\limits_{b=1}^B\vect{h}_{b,u}^{\rm H}[k]\vect{W}_{{\rm RF},b}[k]\vect{H}_{f,b}[k]\vect{v}_{b,u}[k] \right\rvert^2}{\sum\limits_{\substack{q=1,\\q\neq u}}^U\left\lvert \sum\limits_{b=1}^B\vect{h}_{b,u}^{\rm H}[k]\vect{W}_{{\rm RF},b}[k]\vect{H}_{f,b}[k]\vect{v}_{b,q}[k] \right\rvert^2 + \sigma_{u,k}^2}.
\end{equation}

In this paper, assuming imperfect knowledge of $\tilde{\vect{h}}\triangleq\{\vect{h}_{1,1}[k],\ldots,\vect{h}_{B,U}[k]\}$ at the central node being responsible for all BSs hybrid analog and digital beamforming configurations, we focus on solving the following optimization problem: 
\begin{align*}
    \mathcal{OP}: \,&\max_{\widetilde{\vect{W}}_{\rm RF}(\vect{\alpha}_0),\tilde{\vect{v}}} \,\,\, \mathcal{H}_0 \triangleq\mathbb{E}_{\tilde{\vect{h}}}\left\{ \mathcal{R}\left(\widetilde{\vect{W}}_{\rm RF}(\vect{\alpha}_0),\tilde{\vect{v}}\right) \right\} \\
	&\hspace{0.51cm}\text{s.t.} \quad\quad \sum_{k=1}^{K} \left\lVert\vect{V}_b[k]\right\rVert_{\rm F}^2 \leq P_{b}^{\max} \;\forall b,
\end{align*}
where the expectation is taken over the estimation errors of all $BUK$ channel gain vectors in $\tilde{\vect{h}}$, and $P_b^{\max}$ denotes the maximum transmit power per BS. The notation $\widetilde{\vect{W}}_{\rm RF}(\vect{\alpha}_0)$ explicitly indicates that each analog beamforming matrix is parameterized by the resonant strength vector $\vect{\alpha}_0 \triangleq \{\alpha_{0,1,1}, \dots, \alpha_{0,B,N}\}$, which is the tunable parameter of the Lorentzian model in \eqref{eq:Lorentzian}. 
Accordingly, our goal is to maximize the adopted performance metric with respect to the resonant strength factors $\vect{\alpha}_{0,b,n}$ $\forall b,n$, which affect $\widetilde{\vect{W}}_{\rm RF}$. In addition, as will be revealed in the next section, we propose a distributed (i.e., per BS) design solution, according to which the deployed BSs cooperate with minimal information exchange among them, either in coordination by a CPU or in a decentralized manner~\cite{katsanos2026decentralized}. It is also noted that the power constraint in $\mathcal{OP}$ serves as a surrogate for the actual transmit power. This is because the digital precoding variables are defined in terms of currents, i.e., $\mathbf{i}_b = \mathbf{V}_b \mathbf{s}$, as described in Section~\ref{sec:TX_modeling}. Accurately relating this quantity to the consumed power would require accounting for the effective input impedance of the feeding structure, which is beyond the scope of this paper.

%----------------------------------------------------------%
%                      Section Change                      %
%----------------------------------------------------------%
\section{Distributed Sum-Rate Maximization}
To treat $\mathcal{OP}$ in a parallel way, we make use of the structured surrogate function (see, e.g., \cite{liu2019stochastic} and references therein), according to which $\mathcal{OP}$'s objective $\mathcal{H}_0$ is decomposed as $\overline{\mathcal{H}}_0^t(\vect{x},\vect{\xi}^t) = \sum_{b=1}^B\overline{\mathcal{H}}_{0,b}^t(\vect{x}_b,\vect{\xi}^t)$, with: 
\begin{equation} \label{eqn:stoch_surrogate}
\begin{aligned}
    \overline{\mathcal{H}}_{0,b}^t&(\vect{x}_b,\vect{\xi}^t) \triangleq (1-\rho^t)\mathcal{H}_{0,b}^{t-1} + \rho^t\hat{h}_{0,b}(\vect{x}_b,\vect{x}^t,\vect{\xi}^t) \\
    &\quad+ (1-\rho^t)(\vect{f}_{0,b}^{t-1})^{\rm T}(\vect{x}_b - \vect{x}_b^t) - \frac{\tau}{2}\|\vect{x}_b - \vect{x}_b^t \|^2,
\end{aligned}    
\end{equation}
where $\hat{h}_{0,b}(\vect{x}_b,\vect{x}^t,\vect{\xi}^t) \triangleq \frac{1}{B}\mathcal{R}(\vect{x}^t,\vect{\xi}^t) + \nabla_{\vect{x}_b}^{\rm T}\mathcal{R}(\vect{x}^t,\vect{\xi}^t)(\vect{x}_b - \vect{x}_b^t)$. In the previous two expressions, $\vect{x}_b \triangleq [(\vect{v}_{b,u}[k])^{\rm T},\vect{\alpha}_{0,b}^{\rm T}]^{\rm T}$ $\forall u,k$ collects the design variables for the $b$-th BS and $\vect{x}$ represents the entire set of variables at the algorithmic iteration $t\geq0$, while $\rho^t\in(0,1]$ is a properly chosen step-size with $\rho^0=1$, and $\tau>0$ ensures strong concavity of $\overline{\mathcal{H}}_{0,b}^t$. In addition, $\vect{f}_{0,b}^t$ is the accumulation vector that approximates $\nabla\mathbb{E}[\mathcal{R}(\vect{x}^t,\vect{\xi}^t)]$, and is updated recursively as:
\begin{equation} \label{eqn:accum_vector}
    \vect{f}_{0,b}^t \triangleq (1-\rho^t)\vect{f}_{0,b}^{t-1} + \rho^t\nabla_{\vect{x}_b}\mathcal{R}(\vect{x}^t,\vect{\xi}^t),
\end{equation}
where $\vect{f}_{0,b}^{-1} \triangleq \vect{0}$ and $\vect{\xi}^t$ represents a random realization for the channel states $\tilde{\vect{h}}$, while, setting $\mathcal{H}_{0,b}^{-1}=0$, it holds:
\begin{equation}
    \mathcal{H}_{0,b}^t\triangleq\frac{1}{B}\left((1-\rho^t)\mathcal{H}_{0,b}^{t-1} + \rho^t\mathcal{R}(\vect{x}^t,\vect{\xi}^t)\right).
\end{equation}

Apparently, \eqref{eqn:stoch_surrogate} allows us to approximate the highly non-linear and non-concave objective function of $\mathcal{OP}$ with a strongly concave function based on the successive concave approximation framework~\cite{liu2019stochastic}. In the following subsections, we present the distributed solution to $\mathcal{OP}$ with respect to each design parameter in $\vect{x}_b$. After computing the optimal solution $\overline{\vect{x}}_b^t$ for each subproblem, the variables are updated via $\vect{x}_b^{t+1} \triangleq (1-\gamma^t)\vect{x}_b^t + \gamma^t \overline{\vect{x}}_b^t$, and this procedure is repeated until the objective converges within a tolerance $\epsilon > 0$.

\vspace{- 0.2 cm}
\subsection{Linear Precoding Optimization at Each $b$-th BS}
The optimization problem with respect to the linear precoding vectors $\vect{v}_{b,u}[k]$ $\forall u,k$, using \eqref{eqn:stoch_surrogate}, leads to the following simplified subproblem in each $t$-th algorithmic iteration:
\begin{align*}
    \mathcal{OP}_1: \, \overline{\vect{v}}_{b,u}^t[k] \triangleq &\arg\max_{\vect{v}_{b,u}[k]} \,\,\, \overline{\mathcal{H}}_{0,b}^t(\vect{v}_{b,u}[k],\vect{\xi}^t)  \\
	&\hspace{0.70cm}\text{s.t.} \quad \sum_{u=1}^{U}\sum_{k=1}^{K} \left\lVert\vect{v}_{b,u}[k]\right\rVert^2 \leq P_{b}^{\max},
\end{align*}
where the reduced objective function, after excluding the terms irrelevant to $\vect{v}_{b,u}[k]$, is given by the following expression:
\begin{equation*}
    \begin{aligned}
        \overline{\mathcal{H}}_{0,b}^t&(\vect{v}_{b,u}[k],\vect{\xi}^t) = -\frac{\tau}{2}\left\|\vect{v}_{b,u}[k] - \vect{v}_{b,u}^t[k]\right\|^2\\
        &\rho^t\Re\left\{\nabla_{\vect{v}_{b,u[k]}}^{\rm H}\mathcal{R}(\vect{x}^t,\vect{\xi}^t)(\vect{v}_{b,u}[k] - \vect{v}_{b,u}^t[k]) \right\} \\
        &+(1-\rho^t)\Re\left\{ (\vect{f}_{0,\vect{v},b}^{t-1})^{\rm H}(\vect{v}_{b,u}[k] - \vect{v}_{b,u}^t[k])\right\},
    \end{aligned}
\end{equation*}
which is clearly a concave function with respect to $\vect{v}_{b,u}[k]$. 

We first compute the gradient vector $\nabla_{\vect{v}_{b,u[k]}}\mathcal{R}(\vect{x}^t,\vect{\xi}^t)$, whose derivation leads to the following compact expression:
\begin{equation*}
\begin{aligned}
    &\nabla_{\vect{v}_{b,u[k]}}\mathcal{R} = \frac{1}{\ln(2)}\left[\! \frac{1}{\operatorname{MUI}_{u,k}^t}(\tilde{\vect{H}}_{b,u}[k]\vect{v}_{b,u}^t[k] \!+\! r_{u,u,k}^t\tilde{\vect{h}}_{b,u}[k]) \right. \\ 
    &\left. -\!\!\sum\limits_{\substack{q=1,\\q\neq u}}^U \frac{\mathcal{F}_{q,k}^t(1\!+\!\operatorname{SINR}_{q,k}^t)^{-1}}{(\operatorname{MUI}_{q,k}^t)^2}( \tilde{\vect{H}}_{b,q}[k]\vect{v}_{b,u}^t[k] \!+\! r_{q,u,k}^t\tilde{\vect{h}}_{b,q}[k])\! \right]\!\!,
\end{aligned}
\end{equation*}
where the following definitions have been used:
\begin{align}
    \mathcal{F}_{q,k}^t &\triangleq \left\lvert \sum\limits_{b=1}^B\tilde{\vect{h}}_{b,q}^{\rm H}[k]\vect{v}_{b,q}^t[k] \right\rvert^2,\\
    \operatorname{MUI}_{u,k}^t &\triangleq {\sum_{\substack{q=1,q\neq u}}^U\left\lvert \sum\limits_{b=1}^B\tilde{\vect{h}}_{b,u}^{\rm H}[k]\vect{v}_{b,q}^t[k] \right\rvert^2 + \sigma_{u,k}^2},
\end{align}
along with $\tilde{\vect{H}}_{b,u}[k] \triangleq \tilde{\vect{h}}_{b,u}[k]\tilde{\vect{h}}_{b,u}^{\rm H}[k]$, $\tilde{\vect{h}}_{b,u}[k]\triangleq \vect{H}_{f,b}^{\rm H}[k]\vect{W}_{{\rm RF},b}^{\rm H}[k]\vect{h}_{b,u}[k]$, as well as $r_{q,u,k}^t\triangleq \sum_{b'=1,b'\neq b}^B \tilde{\vect{h}}_{b',q}^{\rm H}[k]\vect{v}_{b',u}^t[k]$. Then, after formulating the Lagrangian function for $\mathcal{OP}_1$ and equating its first-order derivative with respect to $\vect{v}_{b,u}[k]$ to the zero vector, the optimum linear precoding vector $\overline{\vect{v}}_{b,u}^t[k]$ can be obtained as:
\begin{equation} \label{eqn:optimal_precoder}
    \overline{\vect{v}}_{b,u}^t[k] = \frac{1}{\tau+2\lambda_b}\!\left( \rho^t\nabla_{\vect{v}_{b,u[k]}}\mathcal{R} \!+\! (1\!-\!\rho^t)\vect{f}_{0,\vect{v},b}^{t-1} + \tau\vect{v}_{b,u}^t[k] \right)\!.
\end{equation}
In this optimal solution, $\lambda_b \ge 0$ denotes the Lagrange multiplier for the convex maximum transmit-power constraint, which satisfies Slater’s condition, and thus can be efficiently obtained via the bisection search method. 

\vspace{- 0.2 cm}
\subsection{Resonance Strength Optimization at Each $b$-th BS}
The resonance strength parameter vector $\vect{\alpha}_{0,b}$ for the PPW-based DMA at each $b$-th BS can be obtained by solving the following simplified unconstrained optimization problem: 
\begin{align*}
    \mathcal{OP}_2: \, \overline{\vect{\alpha}}_{0,b}^t \triangleq \arg\max_{\vect{\alpha}_{0,b}\in\mathbb{R}} \,\,\, \overline{\mathcal{H}}_{0,b}^t(\vect{\alpha}_{0,b},\vect{\xi}^t),
\end{align*}
whose objective function is equal to:
\begin{equation*}
\begin{aligned}
    &\overline{\mathcal{H}}_{0,b}^t (\vect{\alpha}_{0,b},\vect{\xi}^t) = -\frac{\tau}{2}\left\|\vect{\alpha}_{0,b} - \vect{\alpha}_{0,b}^t\right\|^2 \\
    &+\left(\rho^t\nabla_{\vect{\alpha}_{0,b}}\mathcal{R}(\vect{x}^t,\vect{\xi}^t) + (1-\rho^t)(\vect{f}_{0,\vect{\alpha}_{0,b}}^{t-1}\right)^{\rm T}(\vect{\alpha}_{0,b} - \vect{\alpha}_{0,b}^t),
\end{aligned}    
\end{equation*}
which is clearly a concave optimization problem with respect to $\vect{\alpha}_{0,b}$. Hence, its unique optimal solution can be derived by comparing the objective's derivative with the zero vector, yielding the following closed-form expression:
\begin{equation} \label{eqn:optimal_res_strength}
    \overline{\vect{\alpha}}_{0,b}^t = \frac{\rho}{\tau}\nabla_{\vect{\alpha}_{0,b}}\mathcal{R}(\vect{x}^t,\vect{\xi}^t) + \frac{1-\rho^t}{\tau}\vect{f}_{0,\vect{\alpha}_{0,b}}^{t-1} + \vect{\alpha}_{0,b}^t.
\end{equation}

Evidently, according to the above expression for the optimal $\overline{\vect{\alpha}}_{0,b}$ in each algorithmic iteration $t$, it is crucial to compute the gradient $\nabla_{\vect{\alpha}_{0,b}}\mathcal{R}$, whose expression is given by:
\begin{equation}
\begin{aligned}
    \nabla_{\vect{\alpha}_{0,b}}\mathcal{R}&(\vect{x}^t,\vect{\xi}^t) = -\frac{2}{\ln(2)}\sum_{k=1}^K \vect{J}_{\vect{\alpha}_{0,b,k}}^{\rm T} \\
    &\times\Re\left\{ \operatorname{vec}\left(\vect{W}_{{\rm RF},b}^{-\rm H}[k]\widetilde{\vect{U}}_k \vect{W}_{{\rm RF},b}^{-\rm H}[k]\right) \right\},
\end{aligned}
\end{equation}
where the following matrix definitions have been used:
\begin{align}
    \widetilde{\vect{U}}_k &\triangleq \sum_{u=1}^U \frac{(1+\operatorname{SINR}_{u,k}^t)^{-1}}{(\operatorname{MUI}_{u,k}^t)^2}\vect{U}_{u,k}, \\
    \vect{U}_{u,k} &\triangleq \operatorname{MUI}_{u,k}^t\vect{F}_{u,k} - \mathcal{F}_{u,k}^t\vect{M}_{u,k}, \\
    \vect{F}_{u,k} &\triangleq \left(\vect{H}_{b,u}[k]\vect{v}_{b,u}^t[k] \!+\! r_{u,u,k}^t\vect{h}_{b,u}[k]\right)\vect{v}_{b,u}^{t,\rm H}[k]\vect{H}_{f,b}^{\rm H}[k],\!\!\!\!\\
    \vect{M}_{u,k} &\triangleq \vect{M}_{u,k}^1 + \vect{M}_{u,k}^2, \\
    \vect{M}_{u,k}^1 &\triangleq \vect{H}_{b,u}[k]\left( \sum_{q=1,q\neq u}^U\vect{v}_{b,q}^t[k]\vect{v}_{b,q}^{t,\rm H}[k]\right) \vect{H}_{f,b}^{\rm H}[k], \\
    \vect{M}_{u,k}^2 &\triangleq \vect{h}_{b,u}[k]\left( \sum_{q=1,q\neq u}^U r_{q,u,k}^t \vect{v}_{b,q}^{t,\rm H}[k] \right)\vect{H}_{f,b}^{\rm H}[k],
\end{align}
with $\vect{H}_{b,u}[k] \triangleq \vect{h}_{b,u}[k]\vect{h}_{b,u}^{\rm H}[k]$. In addition, the matrix $\vect{J}_{\vect{\alpha}_{0,b,k}}\in\mathbb{R}^{N^2\times N}$ represents the Jacobian matrix defined as $\vect{J}_{\vect{\alpha}_{0,b,k}}\triangleq \frac{\partial \operatorname{vec} \widetilde{\vect{A}}_b^*[k]}{(\partial \vect{\alpha}_{0,b})^{\rm T}}$, with $\widetilde{\vect{A}}_b^*[k] \triangleq (\vect{A}_b^*[k])^{-1}$ (the latter matrix appears in \eqref{eqn:W_RF_definition}).  More specifically, $\vect{J}_{\vect{\alpha}_{0,b,k}}$ is a sparse matrix whose entries result from differentiating each entry of the Lorentzian function with respect to each entry of $\vect{\alpha}_{0,b}$, and it can be computed as follows:
\begin{equation} \label{eqn:Cap_gradient_part2}
[\vect{J}_{\vect{\alpha}_{0,b,k}}]_{m,\ell} =
\begin{cases}
-\frac{f_{0,\ell}^2 + f_k^2}{\left([\vect{\alpha}_{0,b}]_{\ell} f_{0,\ell}\right)^2}, & m = (\ell-1)N + \ell,
 \\[1ex]
\quad\quad\,\,\,\,\,\,\,\, 0, & \text{otherwise},
\end{cases}
\end{equation}
where $m = 1,\dots,N^2$ and $\ell = 1,\dots,N$. The detailed derivation of $\nabla_{\vect{\alpha}_{0,b}}\mathcal{R}$ is omitted due to space limitations. 
\vspace{- 0.2 cm}
\section{Numerical Results and Discussion} \vspace{-0.10cm}
In this section, we present performance evaluation results for the proposed algorithmic framework solving $\mathcal{OP}$. We have considered $B=3$ BSs, equipped with identical PPW-based DMAs of height $h=2.5$~mm filled with air, deployed on the $xy$-plane ($z_{\rm DMA}=0$). The center of each $b$-th BS was placed on the $x$-axis at $x_{b,{\rm DMA}} = L + (b-1)d_{\rm DMA}$, where $L\triangleq\sqrt{N_f}\lambda_c/2$ denotes the array aperture and $\lambda_c$ the carrier wavelength, while $y_{b,{\rm DMA}}=L/2$ $\forall b$. The considered $U=4$ UEs where grouped into two circular clusters of radius $r=2.5$~m. The first group was centered at $x_{\rm cl}^1 = x_{1,\rm DMA} + 0.5(x_{2,\rm DMA} - x_{1,\rm DMA}) - 0.5L$, and the second at $x_{\rm cl}^2 = x_{2,\rm DMA} + 0.5(x_{3,\rm DMA} - x_{2,\rm DMA}) - 0.5L$, both lying in the plane $y=0$ with $z_{\rm cl}^1 = z_{\rm cl}^2 = 80\,$m. The channel simulation followed the free-space formulation of Section~\ref{sec:channel_model}, augmented with wideband Rayleigh fading and distance-dependent pathloss ${\rm PL} \triangleq {\rm PL}_0 R_{b,u}^{-2.5}$, where $R_{b,u}$ is the distance between the $u$-th UE and the center of the $b$-th BS, with ${\rm PL}_0=-30$ dB. In addition, imperfect CSI realizations $\vect{\xi}^t$'s were obtained via $\hat{h} \triangleq \tilde{h} + e$, where $\tilde{h}$ is the actual channel gain and $e\sim\mathcal{CN}(0,\sigma_e^2)$ with $\sigma_e^2 \triangleq \delta|\tilde{h}|^2$ and $\delta=0.2$. 

For all numerical results, we have assumed uniform transmit power $P_b^{\max}=P^{\max}$ $\forall b$ and noise variance $\sigma_{u,k}^2=\sigma^2=-96$~dBm at each UE. The carrier frequency was set as $f_c = 10$ GHz, the system bandwidth as ${\rm BW} = 250$ MHz, and we employed $K=32$ SCs, $N=64$ metamaterial elements, and $N_f=4$ RF chains per PPW-based DMA. The resonance frequency was fixed as $f_{0,n}=f_c+0.5{\rm BW}+10$ MHz for all DMAs, while $f_k\triangleq f_c + (k - \frac{K+1}{2})\frac{\rm{BW}}{K}$ $\forall k=1,\dots,K$. The step-size sequences were chosen as $\rho^t = (t+2)^{-0.60}$ and $\gamma^t=(t+2)^{-0.61}, \tau=10^{-2}$, and $\epsilon=10^{-3}$. For comparison purposes, we have implemented the ``Perfect CSI'' and ``Imperfect CSI'' scenarios, where, in the latter, we considered the same value for $\delta$, but without updating $\vect{\xi}^t$ as in the proposed ``Robust Design'' at each iteration. We have also evaluated the performance for the case of without mutual coupling (``w/o MC'') enforcing $\vect{G}_b[k]=\vect{0}$ $\forall b,k$ in \eqref{eqn:W_RF_definition}. In the results that follow, we have used $100$ independent channel realizations. 
% In Fig.~\ref{fig:Rates_vs_Pb_max}, we illustrate the performance on the average achievable sum rate of the proposed and described benchmark schemes, as a function of the transmit power, where it can be observed that all of them follow a non-decreasing trend as $P^{\max}$ increases. More importantly, it is showcased that the proposed robust design has superior performance to the imperfect CSI cases, while at the same time the former's achievable rates are very close to the case of perfect CSI which serves as an upper bound. In addition, it is observed that designing the DMA by taking into account MC is more beneficial (apart form enhancing accuracy in terms of hardware design), since in this case the spectral efficiency is significantly larger. 

\begin{figure}[!t]
	\centering
	\includegraphics[width=2.75in]{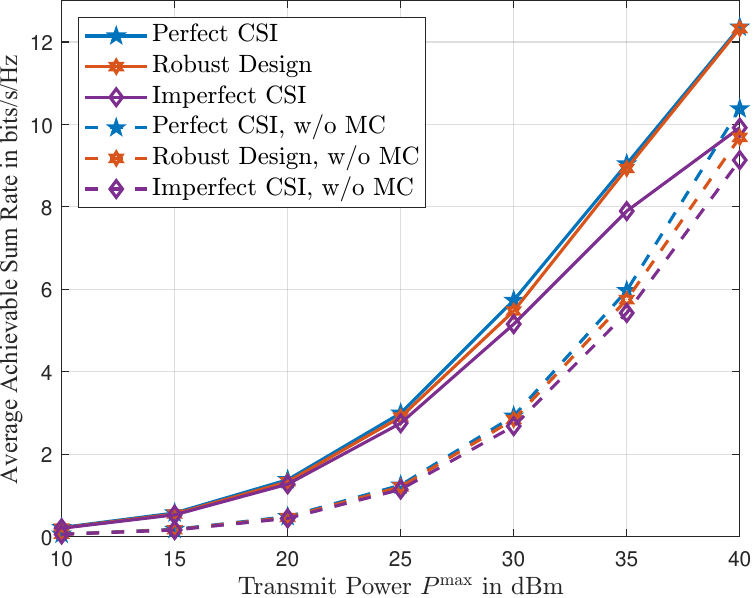}
	\caption{\small{Average achievable sum-rate performance for the considered cell-free OFDM system with $B=3$ PPW-based DMAs, each equipped with $N=64$ metamaterials and $N_f=4$ RF chains. 
    }\vspace{-0.6 cm}}
	\label{fig:Rates_vs_Pb_max}
\end{figure}

Figure~\ref{fig:Rates_vs_Pb_max} depicts the sum rate of the proposed design versus $P_{b}^{\max}$. As shown, all schemes exhibit a non-decreasing behavior as $P^{\max}$ grows. Notably, the proposed robust design consistently outperforms the schemes with imperfect CSI, while attaining achievable rates that remain close to the perfect CSI benchmark, which represents an upper performance bound. Furthermore, accounting for mutual coupling in the DMA design, not only better reflects the hardware characteristics, but also yields a markedly higher spectral efficiency.

%----------------------------------------------------------%
%                      Section Change                      %
%----------------------------------------------------------%
\vspace{- 0.2 cm}
\section{Conclusion} 
In this paper, we studied a cell-free system comprising multiple BSs equipped with identical PPW-based DMAs, and performing downlink OFDM transmissions to serve multiple single-antenna UEs. We focused on distributed beamforming aiming to maximize spectral efficiency under imperfect CSI conditions. Our numerical investigations showcased that the proposed design maintains strong robustness against imperfect CSI, while, taking into account mutual coupling during the DMA design process, yields enhanced performance. 

%----------------------------------------------------------%
%                      REFERENCES                          %
%----------------------------------------------------------%
\vspace{- 0.2 cm}
\bibliographystyle{IEEEtran}
\bibliography{references}
\vspace{- 0.2 cm}
\end{document}